\documentstyle[11pt,epsf]{article}

\def\be{\begin{eqnarray}}
\def\ee{\end{eqnarray}}
\def\ben{\begin{enumerate}}\def\een{\end{enumerate}}

\def\del{\partial}
\def\L{{\cal L}}
\def\E{{\cal E}}
\def\ln{{\rm ln}}
\def\C{\tilde{C}}

\def\prl {Phys. Rev. Lett.}\def\pr{Phys. Rev.}
\def\np{Nucl. Phys.}\def\pl{Phys. Lett.}

\def\roughly#1{\mathrel{\raise.3ex\hbox{$#1$\kern-.75em%
\lower1ex\hbox{$\sim$}}}}

\def\del{\partial}
\renewcommand{\thefootnote}{\fnsymbol{footnote}}
\begin{document}

\setlength{\baselineskip}{24.2pt}

\begin{titlepage}
\hfill{SNUTP 97-157}
\vspace{.3cm}
\begin{center}
\ \\
{\Large \bf Thermodynamic Properties of\\ Effective Chiral
Lagrangians with Brown-Rho Scaling}

\ \\
\vspace{0.2cm}
{Chaejun Song$^{(a,b)}$\footnote{E-mail: chaejun@fire.snu.ac.kr},
Dong-Pil Min$^{(b)}$\footnote{E-mail: dpmin@phya.snu.ac.kr} and
Mannque Rho$^{(a,c)}$\footnote{E-mail: rho@wasa.saclay.cea.fr}}

\vskip 0.5cm

{\it (a) Theory Group, GSI, Planckstr.1, D-64291 Darmstadt, Germany}

{\it (b) Department of Physics and Center for Theoretical Physics,}

{\it Seoul National University, Seoul 151-742, Korea}

{\it (c) Service de Physique Th\'eorique, CEA Saclay,}

{\it F-91191 Gif-sur-Yvette, France}
\vskip 0.3cm
\end{center}
\vskip 0.5cm
\centerline{\bf ABSTRACT}

\noindent We show that effective chiral Lagrangians endowed with 
Brown-Rho  scaling
can be mapped to Landau Fermi-liquid fixed point theory in a way
consistent with general constraints following from thermodynamics.
This provides a unified scheme to treat, starting from normal
nuclear matter, hadronic matter under extreme conditions that is
encountered in relativistic heavy-ion collisions and in the
interior of compact stars.

\vspace{2cm}
\end{titlepage}
\setcounter{footnote}{0}
\renewcommand{\thefootnote}{\arabic{footnote}}
\section{Introduction}
\indent\indent
In a recent publication \cite{SBMR1}, we proposed an effective
chiral Lagrangian whose parameters (masses and coupling constants)
scale in nuclear medium according to the scaling law described by
Brown and Rho \cite{BR91} (referred to in what follows as 
``BR scaling" in short) that, in the mean field approximation, is
to describe the ground state of nuclear matter while when
fluctuated in various flavor directions should enable us to
extrapolate to a density regime beyond the normal matter as well
as to treat excitations above the ground state.
The
simplest such Lagrangian takes the form\footnote{As suggested in
\cite{BR96,PMR}, chiral in-medium Lagrangians can be brought to a
form equivalent to a Walecka-type model. The scalar field appearing
here transforms as a singlet, not as the fourth component of $O(4)$
of the linear sigma model.}
\be
\L&=&\bar{\psi}[\gamma_\mu (i\del^\mu-g_v^\star (\rho )
\omega^\mu )-M^\star (\rho )
+h\phi ]\psi\label{model}\\
& &+\frac12[(\del\phi )^2-m_s^{\star 2}(\rho )\phi^2]-\frac14 F_\omega^2
+\frac12 m_\omega^{\star 2}(\rho )\omega^2\nonumber
\ee
where $\psi$ is the nucleon field, $\omega_\mu$ the isoscalar
vector field, $\phi$ an isoscalar scalar field and the masses with
asterisk are BR-scaling as first introduced in \cite{BR91}. The
scaling behavior of the constant $g_v$ is left arbitrary and the
$h$ is assumed not to scale although it is easy to take into account 
the density dependence if necessary. It was shown
in \cite{SBMR1} that in the mean field, this Lagrangian -- with the
BR scaling suitably implemented -- gives a surprisingly good
description of the ground state with a compression modulus well
within the accepted value $200\sim 300$ MeV.

As it stands, the Lagrangian (\ref{model}) does not look chirally
invariant. This is because we have dropped pion fields which play
no role in the ground state of nuclear matter. In considering
fluctuations around the ground state, they (and other
pseudo-Goldstone fields such as kaons) should be reinstated. The
chiral singlet $\omega$ field and $\phi$ field can be considered as
auxiliary fields brought in from a Lagrangian consisting of
multi-Fermion field operators \cite{PMR} via a Hubbard-Stratonovich
transformation.

The simple Lagrangian (\ref{model}) embodies the effective field
theory of QCD discussed by Furnstahl et al \cite{FTS} anchored on
general considerations of chiral symmetry. As argued in
\cite{SBMR1}, this Lagrangian should be viewed as an effective
Lagrangian that results from two successive renormalization group
``decimations", one leading to a chiral liquid structure \cite{lynn}
at the chiral symmetry scale and the other with respect to the
Fermi surface \cite{shankar}. The advantage of (\ref{model}) is
that it can, on the one hand, be connected to Landau Fermi liquid
fixed point theory of nuclear matter as suggested in
\cite{SBMR1,FR96} and, on the other hand, be extrapolated to the
regime of hadronic matter produced under extreme conditions as
encountered in relativistic heavy ion processes. It would, for
instance, allow one, starting from the ground state of nuclear matter, 
to treat  {\it on the same footing} the dilepton processes
observed in CERES experiments as explained in \cite{LKB} and kaon
production at SIS energy and kaon condensation in dense matter
relevant to the formation of compact stars as discussed in
\cite{LLB}.

In this note, we address an issue which was left unaddressed in
\cite{SBMR1}, namely thermodynamic consistency of the Lagrangian
(\ref{model}) treated in the mean field approximation. For
instance, it is not obvious that the presence of the
density-dependent parameters in the Lagrangian does not spoil the
self-consistency of the model, in particular, energy-momentum
conservation in the medium and also certain relations of
Fermi-liquid structure of the matter\footnote{We thank Bengt Friman
for raising this question.}. The purpose of this paper is to show
that there is no inconsistency in doing a field theory with BR
scaling masses and other parameters.
\section{Implementing BR Scaling in the Lagrangian}
\indent\indent
In \cite{SBMR1}, we have treated the density-dependent masses and
constants as independent of the fields that enter in the
Lagrangian. The Euler-Lagrange equations of motion are then the
same as for the Lagrangian wherein the masses and constants are not
BR-scaling. While this procedure gives correct energy density,
pressure and compression modulus, the energy-momentum conservation
is not automatically assured. In fact, if one were to  compute the pressure 
from the energy-density ${\E}$, one would find that it does not give
$\frac 13 <T_{ii}>$ 
(where $T_{\mu\nu}$ is the conserved energy-momentum tensor and the
bra-ket means the quantity evaluated in the mean-field approximation
as defined before)
unless one drops certain terms
without justification. This suggests that it is incorrect to take
the masses and coupling constants independent of fields in
deriving, by Noether theorem, the energy-momentum tensor. So the
question is: how do we treat the field dependence of the BR scaling
masses and constants?

One possible solution to this problem is as follows. In
\cite{BR91}, the density dependence of the Lagrangian arose as the
``vacuum'' expectation value\footnote{By ``vacuum" we mean the state
of baryon number zero modified from that of true vacuum by the
strong influence of the baryons in the system. See later for more
on this point.} of the scalar field $\chi$ that figures in the QCD
trace anomaly.  It corresponded to the condensate of a quarkonium
component of the scalar $\chi$ with the gluonium component -- which
lies higher than the chiral scale
-- integrated out. It was assumed to scale in dense medium in a skyrmion-type
Lagrangian subject to chiral symmetry. Now in the 
language of a chiral Lagrangian
consisting of the nucleonic matter field $\psi$ with other massive fields
integrated out, this scalar condensate would be some function of
the ``vacuum'' expectation value of $\bar{\psi}\psi$ or
$\bar{\psi}\gamma_0\psi$ coming from
multi-Fermion field operators mentioned above. How these four-Fermi
and higher-Fermi field terms can lead to BR scaling in the
framework of chiral perturbation theory was discussed in
\cite{PMR}. We shall follow this strategy in this paper leaving
other possibilities (such as dependence on the mean fields of the massive
mesons) for later investigation. For this, it is
convenient to define
\be
\check\rho u^\mu \equiv \bar\psi\gamma^\mu\psi
\ee
with unit fluid
4-velocity $u^\mu =\frac{1}{\sqrt{1-\vec{v}^2}}(1,\vec{v})
=\frac{1}{\sqrt{\rho^2-\vec{j}^2}}(\rho ,\vec{j})$
with the baryon current density $\vec{j}= <\bar\psi\vec\gamma\psi
>$ and the baryon number density $\rho =<\psi^\dagger\psi >
=\sum_i\rho_i$. We will take $\rho_i$ to be given by the Fermi distribution
function,  $\rho_i=\theta (k_F-|\vec{k}_i|)$ at $T=0$. We should
replace $\rho$ in (\ref{model}) by $\check\rho$ for consistency of
the model. The definition of $\check\rho$ makes our Lagrangian
Lorentz-invariant which will later turn out to be useful in
deriving relativistic Landau formulas. With this, the
Euler-Lagrange equation of motion (EOM) for the nucleon field is
\be
\frac{\delta\L}{\delta\bar\psi}&=&
\frac{\del\L}{\del\bar\psi}+\frac{\del\L}{\del\check\rho}\frac{\del\check\rho}
{\del\bar\psi} \nonumber\\
&=&[i\gamma^\mu (\del_\mu +ig_v^\star\omega_\mu -iu_\mu\check\Sigma )
-M^\star +h\phi ]\psi=0\label{fer}
\ee
with
\be
\check\Sigma &=&\frac{\del\L}{\del\check\rho}\nonumber\\
&=&m_\omega^\star\omega^2\frac{\del m_\omega^\star}{\del\check\rho}
-m_s^\star\phi^2\frac{\del m_s^\star}{\del\check\rho}
-\bar\psi\omega^\mu\gamma_\mu\psi\frac{\del g_v^\star}{\del\check\rho}
-\bar\psi\psi\frac{\del M^\star}{\del\check\rho}.
\ee
This additional term which may be related to what is referred to in
many-body theory as ``rearrangement terms" plays a crucial role in
what follows. The EOM's for the bosonic fields are
\be
(\del^\mu\del_\mu +m_s^{\star 2})\phi &=& h\bar{\psi}\psi\label{ph} \\
\del_\nu F_\omega^{\nu\mu}+m_\omega^{\star 2}\omega^\mu &=& g_v^\star
\bar{\psi} \gamma^\mu \psi.
\ee
\section{Equation of State (EOS)}
\indent\indent
We start with the conserved canonical energy-momentum tensor
constructed a la Noether from the Lagrangian (\ref{model}):
\be
T^{\mu\nu}&=&i\bar\psi\gamma^\mu\del^\nu\psi +\del^\mu\phi\del^\nu\phi
-\del^\mu\omega_\lambda\del^\nu\omega^\lambda\nonumber\\
& &-\frac12 [(\del\phi )^2
-m_s^{\star 2}\phi^2-(\del\omega )^2+m_\omega^{\star 2}\omega^2
-2\check\Sigma\bar\psi\gamma\cdot u\psi ]g^{\mu\nu}.
\label{emtensor}
\ee
We shall compute thermodynamic quantities from (\ref{emtensor})
using the mean field approximation which amounts to taking
\be
\psi &=&\frac{1}{\sqrt{V}}\sum_i
a_i\sqrt{\frac{E_{\kappa_i}+m_L^\star}{2E_{\kappa_i}}}
\left(\begin{array}{c}
\chi\\
\frac{\vec\sigma\cdot\vec{\kappa_i}}{E_{\kappa_i}+m_L^\star}\chi
\end{array}\right)
\exp{(i\vec{\kappa}_i\cdot\vec{x}-i(g_v^\star\omega_0-u_0\Sigma+E_i) t)}
\label{fermi}\\
h\phi &=&C_h^2<\bar\psi\psi >=C_h^2\sum_i\rho_i\frac{m_L^\star }
{\sqrt{\vec{\kappa}^2_i+m_L^{\star 2}}}\label{phi}\\
g_v^\star\omega_0 &=& C_v^2\rho =C_v^2\sum_i\rho_i\label{om0}\\
g_v^\star\vec\omega &=&C_v^2\vec{j}=C_v^2\sum_i\rho_i\frac{\vec{\kappa}_i}
{\sqrt{\vec{\kappa}^2_i+m_L^{\star 2}}}\label{omega}
\ee
where $a_i$ is the annihilation operator of the nucleon $i$,
with $\rho_i =<a^\dagger_i a_i>$, and $\Sigma =\langle\check\Sigma\rangle$,
$E_{\kappa_i} =\sqrt{\vec{\kappa}^2_i+m_L^{\star 2}}$ with
$m_L^\star\equiv M^\star -h\phi$,
$\chi$ is the spinor and $\vec\sigma$ is the Pauli matrix. We have defined
\be
C_v(\check\rho )&\equiv&
\frac{g_v^\star (\check\rho )}{m_\omega^\star (\check\rho )}\\
C_h(\check\rho )&\equiv&\frac{h}{m_s^\star (\check\rho)}\\
\C_h(\check\rho)&\equiv&\frac{1}{C_h(\check\rho )}\\
\vec{\kappa}&\equiv&\vec{k}-C_v^2\vec{j}+\vec{u}\Sigma .
\ee
In this approximation, the energy density is
\be
\E &=&<T^{00}>\nonumber\\
&=&<i\bar\psi\gamma^0\del^0\psi+\frac12 m_s^{\star 2}\phi^2
-\frac12 m_\omega^{\star 2}\omega^2+\check\Sigma\bar\psi\gamma^\mu u_\mu\psi >
\nonumber\\
&=& \frac12 C_v^2(\rho^2+\vec{j}^2)
+\frac12 \C_h^2(m_L^\star -M^{\star })^2
+\sum_l\rho_l\sqrt{\vec{\kappa}^2_l+m_L^{\star 2}}-
\Sigma\vec{u}\cdot\vec{j}\label{edensity}.
\ee
Note that the $\Sigma$-dependent terms cancel out in the comoving
frame ($\vec{v}
=0$), so that the resulting
energy-density is identical to what one would obtain from the
Lagrangian in the mean field with the density-dependent parameters
{\it taken as field-independent quantities} as in \cite{SBMR1}.

Given the energy density (\ref{edensity}), the pressure can be
calculated by (at $T=0$)
\be
p&=&-\frac{\del E}{\del V}=\rho^2\frac{\del\E
/\rho}{\del\rho}=\mu\rho -\E\nonumber\\
&=& \frac12 C_v^2(\rho )\rho^2 -\Sigma_0 \rho
-\frac12 \C_h^2(\rho )(m_L^\star -M^{\star }(\rho ))^2\nonumber\\
& &-\frac{\gamma}{2\pi^2}\left( E_F(\frac{m_L^{\star 2}}{8}k_F
-\frac{1}{12}k_F^3)-\frac{m_L^{\star 4}}{8}\ln [(k_F+
E_F)/m_L^\star ]\right)\label{P}
\ee
where $\mu$ is the chemical potential --
the first derivative of the energy density with respect
to $\rho$ in the comoving frame ($\vec{v}=0$):
\be
\mu\equiv \frac{\del}{\del\rho}\E|_{\vec{v}=0}
=C_v^2\rho +E_F-\Sigma_0\label{mu}
\ee
with $E_F=\sqrt{k_F^2+m_L^{\star 2}}$ and $\Sigma_0=\langle\check\Sigma
\rangle_{\vec{v}=0}$. To check that this is consistent, we calculate the
pressure from the energy-momentum tensor (\ref{emtensor}) in the
mean field at $T=0$:
\be
p_t&\equiv&\frac13 <T_{ii}>_{\vec{v}=0}\nonumber\\ &=&\frac13
\langle i\bar\psi\gamma^i\del^i\psi
-\frac{1}{2}(m_\omega^{\star 2}\omega^2-m_s^{\star 2}\phi^2
-2\check\Sigma\psi^\dagger\psi )g^{ii}\rangle_{\vec{v}=0}
\nonumber\\
&=&\frac12 C_v^2(\rho )\rho^2
-\frac12 \C_h^2(\rho )(m_L^\star -M^{\star }(\rho ))^2-\Sigma_0 \rho\nonumber\\
&&-\frac{\gamma}{2\pi^2}\left( E_F(\frac{m_L^{\star 2}}{8}k_F
-\frac{1}{12}k_F^3)-\frac{m_L^{\star 4}}{8}\ln [(k_F+
E_F)/m_L^\star ]\right).
\ee
This agrees with (\ref{P}). Thus our EOS conserves energy and
momentum.
%
\section{Landau Fermi-Liquid Parameters}
\indent\indent
The next issue we address is the connection between the mean-field
theory of the chiral Lagrangian (\ref{model}) and Landau's
Fermi-liquid fixed point theory as formulated in \cite{SBMR1,FR96}.
As far as we know, this connection is the only means available to
implement chiral symmetry of QCD in dense matter based on effective
field theory. For this, we shall follow closely Matsui's analysis
of Walecka mean field model \cite{matsui} exploiting the similarity
of our model to the latter.
\subsection{\it Quasiparticle interactions}
The quasiparticle energy $\varepsilon_i$ and quasiparticle
interaction  $f_{ij}$ are, respectively, given by first and second
derivatives with respect to $\rho_i$:
\be
\varepsilon_i=&\frac{\del\E}{\del\rho_i}, \ \ \ f_{ij}=\frac{\del\varepsilon_i}{\del\rho_j}.
\ee
A straightforward calculation gives
\be
\varepsilon_i
&=&C_v^2\rho +\sqrt{\vec{\kappa}_i^2+m_L^{\star 2}}
+C_v\rho^2\frac{\del C_v}{\del\rho_i}
-C_v^2\vec{j}^2\frac{\del C_v}{\del\rho_i}
\nonumber\\
& &+\C_h(m_L^\star-M^\star )^2\frac{\del \C_h}{\del\rho_i}
-\C_h^2(m_L^\star-M^\star )\frac{\del M^\star}{\del\rho_i}
-\Sigma\vec{u}\cdot\frac{\del\vec{j}}{\del\rho_i}\label{quasienergy}
\ee
and
\be
f_{ij}&=&\frac{\del\varepsilon_i}{\del\rho_j}|_{\vec{j}=\vec{v}=0}\nonumber\\
&=&C_v^2+4C_v\rho\frac{\del C_v}{\del\rho}
+\frac{m_L^\star}{E_i}\frac{\del m_L^\star}{\del\rho_j}
+\rho^2[(\frac{\del C_v}{\del\rho})^2+C_v\frac{\del^2C_v}{\del\rho^2}]
\nonumber\\
& &+(m_L^\star -M^\star )^2[(\frac{\del \C_h}{\del\rho})^2
+\C_h\frac{\del^2\C_h}{\del\rho^2}]
+2\C_h\frac{\del \C_h}{\del\rho}(m_L^\star -M^\star )
\frac{\del}{\del\rho_j}(m_L^\star -M^\star )\nonumber\\
& &-2\C_h\frac{\del \C_h}{\del\rho}(m_L^\star -M^\star )
\frac{\del M^\star}{\del\rho}-\C_h^2\frac{\del M^\star}{\del\rho}
\frac{\del}{\del\rho_j}(m_L^\star -M^\star )
-\C_h^2(m_L^\star -M^\star )\frac{\del^2 M^\star}{\del\rho^2}\nonumber\\
& &-(C_v^2-\frac{\Sigma_0}{\rho} )
\frac{\vec{k_i}}{E_i}\cdot\frac{\del\vec{j}}{\del\rho_j}.
\label{fij}\ee
with $E_i=\sqrt{\vec{k}_i^2+m_L^{\star 2}}$. Note that $C_v$,
$\C_h$, and $M^\star$ are functions of $\langle\check\rho\rangle
=u_0\rho-\vec{u}\cdot\vec{j}$ in the mean field approximation. In
arriving at (\ref{fij}), we have used the observation that in the limit 
$\vec{j}\rightarrow 0$, we have 
\be
\frac{\del u_0}{\del\rho_i}
&\rightarrow& 0,\nonumber\\
\frac{\del^2u_0}{\del\rho_i\del\rho_j}
&\rightarrow&\frac{1}{\rho^2}
\frac{\del\vec{j}}{\del\rho_i}\cdot\frac{\del\vec{j}}{\del\rho_j},\nonumber\\
\frac{\del\vec{u}}{\del\rho_i}
&\rightarrow&
\frac{1}{\rho}\frac{\del\vec{j}}{\del\rho_i},\nonumber\\
\frac{\del\langle\check\rho\rangle}{\del\rho_i}
&\rightarrow& 1,\nonumber\\
\frac{\del^2\langle\check\rho\rangle}{\del\rho_i\del\rho_j}
&\rightarrow &-\frac{1}{\rho}
\frac{\del\vec{j}}{\del\rho_i}\cdot\frac{\del\vec{j}}{\del\rho_j}\nonumber
\ee
and that if $f$ is taken to be a function of the expectation value of
$\check\rho$, then as $\vec{j}\rightarrow 0$, we have
\be
\frac{\del f}{\del\rho_i}&=&\frac{\del f}{\del \langle\check\rho\rangle}
\frac{\del\langle\check\rho\rangle}{\del\rho_i}
\rightarrow \frac{\del f}{\del \rho}\\
\frac{\del^2f}{\del\rho_i\del\rho_j}
&=&\frac{\del^2f}{\del\langle\check\rho\rangle^2}
\frac{\del\langle\check\rho\rangle}{\del\rho_i}\cdot
\frac{\del\langle\check\rho\rangle}{\del\rho_j}
+\frac{\del f}{\del\langle\check\rho\rangle}
\frac{\del^2\langle\check\rho\rangle}{\del\rho_i\del\rho_j}\nonumber\\
&\rightarrow&\frac{\del^2f}{\del\rho^2}
-\frac{1}{\rho}\frac{\del f}{\del\rho}
\frac{\del\vec{j}}{\del\rho_i}\cdot\frac{\del\vec{j}}{\del\rho_j}.
\ee
In the absence of the baryon current, $\vec{j}=0$, the quantities
$\frac{\del m_L^\star}{\del\rho_j}$ and
$\frac{\del\vec{j}}{\del\rho_j}$  simplify to
\be
\frac{\del m_L^*}{\del\rho_j}=\frac{
\frac{\del M^\star}{\del\rho}-2C_h\frac{\del C_h}{\del\rho}
\sum_l\rho_l\frac{m_L^\star}{E_l}
-C_h^2\frac{m_L^\star}{E_j}}
{1+C_h^2\sum_l\rho_l\frac{\vec{k}_l^2} {E_l^{3/2}}}
\ee
and
\be
\frac{\del\vec{j}}{\del\rho_j}=
\frac{\frac{\vec{k}_j}{E_j}}
{1+(C_v^2-\frac{\Sigma_0}{\rho})\sum_l\rho_l\frac{
\frac{2}{3}\vec{k}_l^2+m_L^{\star 2}}{E_l^{3/2}}}.\label{jeq}
\ee
Writing in the standard way
\be
f_l=(2l+1)\int\frac{d\Omega}{4\pi}P_l(\frac{\vec{k}_i\cdot\vec{k}_j}{k_F^2})
f_{ij}(|\vec{k}_i|=|\vec{k}_j|=k_F),
\ee
we see that the last term in (\ref{fij}) contributes to $f_1$ and
the sum of the rest at the Fermi surface (i.e. $|\vec{k}_j|=k_F$)
to $f_0$. So
\be
F_0&\equiv&\frac{\gamma k_FE_F}{2\pi^2}f_0=\frac{3E_F}{k_F}\rho
f_0\nonumber\\ &=&\frac{3E_F}{k_F}\rho [ C_v^2+4C_v\rho\frac{\del
C_v}{\del\rho} +\frac{m_L^\star}{E_F}\frac{\del
m_L^\star}{\del\rho_j} +\rho^2[(\frac{\del
C_v}{\del\rho})^2+C_v\frac{\del^2C_v}{\del\rho^2}]
\label{f0}\\
& &+(m_L^\star -M^\star )^2[(\frac{\del \C_h}{\del\rho})^2
+\C_h\frac{\del^2\C_h}{\del\rho^2}]
+2\C_h\frac{\del \C_h}{\del\rho}(m_L^\star -M^\star )
\frac{\del}{\del\rho_j}(m_L^\star -M^\star )\nonumber\\
& &-2\C_h\frac{\del \C_h}{\del\rho}(m_L^\star -M^\star )
\frac{\del M^\star}{\del\rho}-\C_h^2\frac{\del M^\star}{\del\rho}
\frac{\del}{\del\rho_j}(m_L^\star -M^\star )
-\C_h^2(m_L^\star -M^\star )\frac{\del^2 M^\star}{\del\rho^2}\nonumber
\ee
and
\be
F_1\equiv\frac{\gamma k_FE_F}{2\pi^2}f_1
=-\frac{3(C_v^2-\frac{\Sigma_0}{\rho} )\rho}
{E_F+(C_v^2-\frac{\Sigma_0}{\rho} )\rho}.
\label{f1}
\ee
\subsection{\it Compression modulus and $F_0$}

The compression modulus K defined by
\be
K\equiv 9\rho\frac{\del^2\E (\vec{j}=0)}{\del\rho^2}
\ee
comes out to be
\be
K&=&\frac{3k_F^2}{E_F}+9\rho [C_v^2
+4C_v\rho\frac{\del C_v}{\del\rho}
+\frac{m_L^*}{E_F}\frac{\del m_L^\star}{\del\rho}
+\rho^2\{(\frac{\del C_v}{\del\rho})^2
+C_v\frac{\del^2C_v}{\del\rho^2}\}
\label{modul}\\
& &+(m_L^\star -M^\star )^2\{(\frac{\del \C_h}{\del\rho})^2
+\C_h\frac{\del^2\C_h}{\del\rho^2}\}
+2\C_h\frac{\del \C_h}{\del\rho}
(m_L^\star -M^\star )\frac{\del}{\del\rho}(m_L^\star -M^\star )
\nonumber\\
& &-2\C_h\frac{\del \C_h}{\del\rho}(m_L^\star -M^\star )
\frac{\del M^\star}{\del\rho}-\C_h^2\frac{\del M^\star}{\del\rho}
\frac{\del}{\del\rho}(m_L^\star -M^\star )
-\C_h^2(m_L^\star -M^\star )\frac{\del^2 M^\star}{\del\rho^2}]
\nonumber\ee
Comparing (\ref{f0}) and ({\ref{modul}), we verify that our model
satisfies the relativistic Landau Fermi-liquid formula for the
compression modulus \cite{baymchin};
\be
K=\frac{3k_F^2}{E_F}(1+F_0).\label{K}
\ee
\subsection{\it First sound velocity}

The first sound velocity $c_1$ in the relativistic case is defined
by
\be
c_1^2&\equiv&\frac{\del p}{\del\E}\nonumber\\
&=&\frac{\del}{\del\E}\left( \rho^2\frac{\del\E/\rho}{\del\rho}\right)
\nonumber\\
&=&\frac{\del\rho}{\del\E}\frac{\del}{\del\rho}(\mu\rho -\E)\nonumber\\
&=&\frac{K}{9\mu}.
\ee
{}From (\ref{K}), we have that \cite{baymchin}
\be
c_1=v_F\sqrt{\frac{E_F}{3\mu}(1+F_0)}.
\ee
This is of course satisfied in our model.
\subsection{\it Relativistic Landau effective mass}

Baym and Chin have shown \cite{baymchin}
that the relativistic Landau liquid satisfies
the mass relation
\be
k_F\left(\frac{\del k_i}{\del\varepsilon_i}\right)_{k=k_F,\vec{v}=0}
=\mu (1+F_1/3).\label{mass}
\ee
In our model $k_F\left(\frac{\del
k_i}{\del\varepsilon_i}\right)_{k=k_F,\vec{v}=0}=E_F$. One can see
from equations (\ref{mu}) and (\ref{f1}) that (\ref{mass}) is
satisfied exactly in our model.
\section{Discussions}
\indent\indent
We showed that a simple effective chiral Lagrangian with BR
scaling parameters is thermodynamically consistent, a point which
is important for studying nuclear matter under extreme conditions.
It is clear however that this does not require that the masses
appearing in the Lagrangian scale according to BR scaling only.
What is shown in this paper is that masses and coupling constants
could depend on density without getting into inconsistency with
general constraints of chiral Lagrangian field theory. This point is
important for applying (\ref{model}) to the density regime
$\rho\sim 3\rho_0$ appropriate for the CERES dilepton experiments
and also kaon production at GSI where deviation from the simple BR
scaling of \cite{BR91} might occur.

The crucial question is really how to understand the scaling masses
and constants as one varies temperature and density as considered
in \cite{BR91}. If one takes the basic assumption of
\cite{SBMR1,FR96} that the chiral Lagrangian in the mean field with
BR scaling parameters corresponds to Landau's Fermi-liquid fixed
point theory, then one should consider first fixing the Fermi
momentum $k_F$ and let renormalization group flow come to the fixed
points of the effective mass $M^\star$ for the nucleon and Landau
parameters ${\cal F}$ \cite{shankar}. In this case, the scaling
quantities would seem to be dependent upon $\Lambda/k_F$, not on
the fields entering into the effective Lagrangian. This paper
however shows that if one wants to approach the Fermi-liquid fixed
point theory {\it starting from an effective chiral Lagrangian of
QCD}, it is necessary to take into account the fact that the
scaling arises from the effect of multi-Fermi interactions figuring
in chiral Lagrangians as implied by chiral perturbation theory
described in \cite{PMR}. This is probably due to the fact that we
are dealing with two-stage ``decimations'' in the present problem --
with the Fermi surface formed from a chiral Lagrangian as a
nontopological soliton (i.e., ``chiral liquid" \cite{lynn}) -- in contrast to
condensed matter situations where one starts {\it ab initio} with
the Fermi surface without worrying about how the Fermi surface is
formed. Our result suggests that there will be a duality in
describing processes manifesting the scaling behavior. In other
words, the change of ``vacuum" by density exploited in \cite{BR91}
could equally be represented by a certain (possibly infinite) set
of interactions among hadrons -- e.g., four-Fermi and higher-Fermi
terms in chiral Lagrangians -- canonically taken into account in
many-body theories starting from the usual matter-free vacuum. A
notable evidence may be found in the two plausible explanations of
the low-mass enhancement in CERES dilepton yields in terms of
scaling vector-meson masses \cite{LKB} and in terms of hadronic
interactions giving rise to increased widths \cite{wambach}.

How to go from one decimation to the next in hadronic physics
remains an open problem as stressed in \cite{SBMR1}.

\subsection*{Acknowledgments}

We have benefited from  valuable discussions with Gerry Brown and Bengt
Friman. Part of this work was done while two of us (M.R. and C.S.) were
visiting the Theory Group of GSI whose hospitality is gratefully
acknowledged. The work of M.R. at GSI was supported by a
Franco-German Humboldt Research Prize and the work of C.S. and D.P.M. 
in part by KOSEF through the Center for Theoretical Physics of Seoul National
University  and in part by Korea Ministry of Education 
(Grant No. BSRI97-2418).


\newpage
\begin{thebibliography}{99}
\bibitem{SBMR1} C. Song, G. E. Brown, D.-P. Min, and M. Rho,
Phys. Rev. C {\bf 56} (1997) 2244
\bibitem{BR91} G. E. Brown and M. Rho, Phys. Rev. Lett. {\bf 66} (1991) 2720
\bibitem{BR96} G.E. Brown and M. Rho, \np \ {\bf A596} (1996) 503
\bibitem{PMR} T.-S. Park, D.-P. Min and M. Rho, \np \ {\bf A596} (1996) 515;
G. Gelmini and R. Ritzi, \pl \ {\bf B357} (1995) 431
\bibitem{FTS} R.J. Furnstahl, H.-B. Tang and B.D. Serot, \pr \ C {\bf 52} (1995)
1368
\bibitem{lynn} B.W. Lynn, \np \ {\bf B402} (1993) 281
\bibitem{shankar} J. Polchinski, {\it Recent Directions in Particle
Theory: From Superstrings and Black Holes to the Standard Model (TASI-92)}, 
ed. J. Harvey and J. Polchinski (World Scientific, Singapore, 1994)
235; R. Shankar, Rev. Mod. Phys. {\bf 66} (1994) 129; T. Chen, J.
Fr\"olich and M. Seifert, cond-mat/9508063
\bibitem{FR96} B. Friman and M. Rho, Nucl.
Phys. {\bf A606} (1996) 303
\bibitem{LKB} G.Q. Li, C.M. Ko, and G.E. Brown, \prl \ {\bf 75}(1995) 4007;
\np \ {\bf A606}(1996) 568
\bibitem{LLB} G.Q. Li, C.-H. Lee and G.E. Brown, \prl (in press) and
nucl-th/9711002
\bibitem{matsui} T. Matsui, \np \ {\bf A370} (1981) 365
\bibitem{baymchin} G. Baym and S. Chin, \np \ {\bf A262}(1976) 527
\bibitem{wambach} W. Cassing, E.L. Bratkovskaya, R. Rapp and J. Wambach,
nucl-th/9708020
\end{thebibliography}
\end{document}